\documentclass[twocolumn,showpacs,preprintnumbers,amsmath,amssymb]{revtex4}
\usepackage{graphicx}
\usepackage{dcolumn}
\usepackage{amsmath}
\usepackage{bm}
\def\bea {\begin{eqnarray}}
\def\eea {\end{eqnarray}}

\def\be {\begin{equation}}
\def\ee {\end{equation}}

\begin{document}

\title{Anomalous features of particle production in high-multiplicity events of $pp$ collisions at the LHC energies}
\author{\it Samrangy Sadhu}
\author{\it Premomoy Ghosh}
\email{prem@vecc.gov.in}
\address{Variable Energy Cyclotron Centre, HBNI, 1/AF Bidhan Nagar, Kolkata 700 064, India}
        
\date{\today}
\begin{abstract}
Prevalent models of multi-particle production in relativistic $pp$ collisions at pre-LHC energies, fail to provide convincing explanations to certain significant features
of the final-state charged particles in high-multiplicity $pp$ events at the LHC. This article presents a study, contrasting those features, usually interpreted as collective
behaviour of particle production in relativistic heavy-ion collisions, in the framework of hydrodynamic EPOS3 model, emphasizing quantitative comparison between the 
data and the model-based simulation in the same kinematic ranges, for better cognizance of the data. The work reveals quantitative mismatch between the data and 
the model.\\
\end{abstract}

\maketitle

\section{Introduction}
\label{}
The experiments at the Relativistic Heavy Ion Collider (RHIC), through satisfactory description of the flow-like behaviour of particle production in relativistic heavy-ion collisions 
by relativistic hydrodynamics \cite {ref01}, established \cite {ref02, ref03, ref04, ref05} formation of collective medium in such collisions. Suppression of high-$p_{T}$ particles, 
revealed in terms of relative yield of charged particles in heavy-ion collisions compared to pp collisions at the same centre-of mass energy, identified the collective medium as 
thermalized medium of Quark-Gluon Plasma (QGP) \cite {ref06, ref07}. Reconfirming the formation of QGP with heavier ions at higher collisional energies, the experiments at 
the Large Hadron Collider (LHC), reported flow-like features of particle productions, in high-multiplicity events of $pp$ collisions, in contrast to the present understanding of 
particle production in relativistic $pp$ collisions, based on the data at the pre-LHC energy range. It is the so-called ``ridge" structure in the long range near side angular correlations 
in high-multiplicity events of proton-proton collisions at $\sqrt s$ = 7 TeV that first reported \cite {ref08} the anomaly in particle productions in $pp$ collisions at the LHC. Further 
experimental studies \cite {ref09, ref10, ref11} at $\sqrt s$ = 7 and 13 TeV confirmed the anomaly by extracting elliptic flow coefficient, $v_{2}$, mass ordering of $v_{2}(p_{T})$ for 
identified charged particles \cite {ref11} and $p_{T}$-dependence of $v_{2}$ \cite {ref09}. Similar unexpected features of particle production have also been observed in $pPb$ 
collisions at $\sqrt s_{NN}$ = 5.02 TeV \cite {ref12, ref13, ref14, ref15} at the LHC. Subsequently, the RHIC data of $dAu$ collisions uphold \cite {ref16} the LHC observation of 
collective property in particle production from small systems. The observation of elliptic and triangular flow patterns of charged particles produced in pAu, dAu and $^{3}$HeAu 
collisions at $\sqrt s_{NN}$ = 200 GeV have been reported \cite {ref17} to be described best by hydrodynamic models, which include the formation of a short-lived QGP droplet. 
Theoretically, the appearance of ``ridge" structure in high-multiplicity pp events at 7 TeV has been shown \cite {ref18} to be expected in hydrodynamic approach, based on flux tube initial 
conditions. The observation of the collective behaviour of particle production in high-multiplicity pp events was further corroborated with the strong transverse radial flow extracted 
\cite {ref19} from satisfactory description of identified charged particle yields \cite {ref20} from high-multiplicity pp events by the hydrodynamics-motivated Boltzmann - Gibbs 
blast-wave (BGBW) model \cite {ref21}. Several other theoretical and phenomenological studies \cite {ref22, ref23, ref24, ref25, ref26} indicate the possibility of a hydro-like collective 
medium in high-multiplicity pp events. In spite of all these studies, suggesting collective behaviour of particle production  in high-multiplicity events of small systems, the idea 
of connecting these anomalous features with the established hydro-like features of particle production from the QGP-like thermalized medium is not unanimously acceptable yet 
because of non-observance of the signal of suppression of high-$p_{T}$ particles or any other compelling signal of formation of the medium. Further, while several models, in different 
approaches, qualitatively describe the data, it becomes difficult to conclude on physics origin of particle production. The source of the ``collective" or the flow-like features 
in the small systems thus remains ambiguous and invite thorough studies involving quantitative comparison of the data in the light of several existing models of particle productions in 
relativistic $pp$ collisions. \\

This article presents a comprehensive study on high-multiplicity $pp$ events at $\sqrt s$ = 7 and 13 TeV, comparing the data with EPOS3 \cite {ref27} generated simulated events, 
with the aim to understand the anomalous features of particle production in $pp$ collisions at the LHC. The EPOS3 simulation generates events with and without hydrodynamical 
evolution. The hydrodynamic EPOS3 event generator follows similar particle production mechanism in $proton-proton$, $proton-nucleus$ and $nucleus-nucleus$ collisions and thus 
become a suitable testing ground for understanding the observed flow-like features in the high-multiplicity $pp$ and $pPb$ events at the LHC in comparison with the well studied 
collective phenomena in relativistic $nucleus-nucleus$ collisions. In the following section, we describe the event generator, in brief. We analyse the EPOS3 generated events and 
present the results in Section III, in terms of the experimental observables which exhibit the flow-like effects in high-multiplicity $pp$ events, namely i) the two-particle azimuthal 
correlations among charged particles, ii) the blast-wave description of identified charged particles iii) the mean transverse momentum ($\langle p_{T} \rangle$) as a function of mean 
charged multiplicity ($\langle N_{ch} \rangle$) and iv) the inverse slope parameter of transverse mass ($m_{T}$)-distribution. All these observables are related to the transverse
momentum of produced particles, collectively presented as the inclusive charged particle transverse momentum spectra. Before carrying out differential analysis of the generated 
events it terms of these observables for a closer comparison with the data on an equal basis, the measured inclusive charged particle transverse momentum spectra for $pp$ collisions 
at $\sqrt s$ = 7 and 13 TeV are matched with those obtained from simulated events to ensure that the simulation code is, at least, minimally tuned. We summarise and conclude 
in Section IV. \\

\begin{center}
\begin{figure}
\includegraphics[scale=0.45]{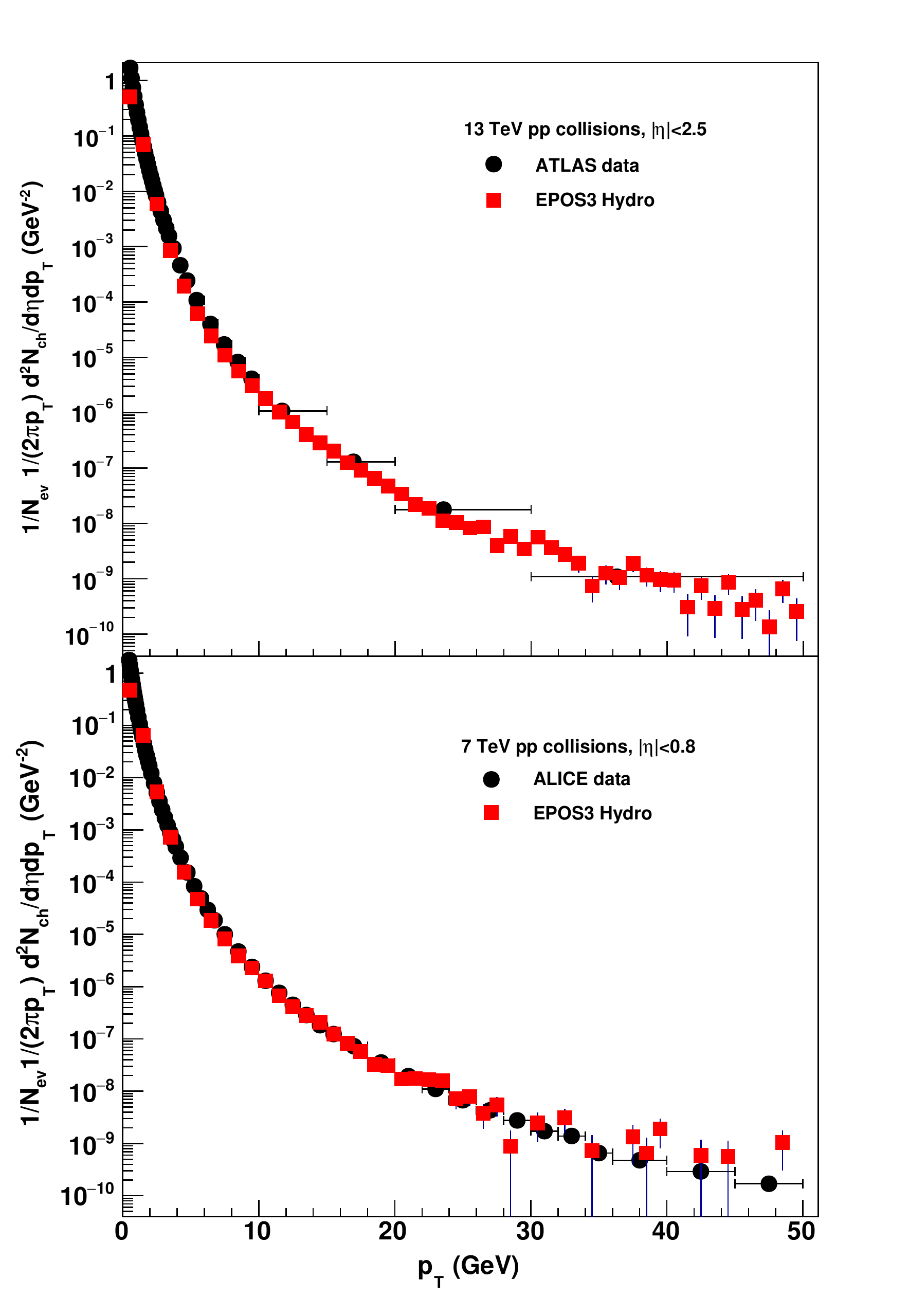}
\caption{Inclusive transverse momentum spectra of charged particles from generated minimum-bias events of $pp$ collisions at $\sqrt s$ = 7 and 13 TeV  from EPOS3 event generator, 
with hydrodynamic calculations, are compared with data, as measured by ALICE \cite {ref28} (lower panel) and ATLAS \cite {ref29} (upper panel) experiments, respectively. }
\label{fig:incl_spectra} 
\end{figure}
\end{center}

\section{Event Generator}
\label{}
The detail of the EPOS3 model parameters are discussed in the ref.~\cite{ref27}. The model is known as Parton-based Gribov-Regge theory. In the Gribov-Regge multiple scattering 
framework, an individual scattering, referred to as Pomeron, gives rise to parton ladder that consists of a hard pQCD scattering along with initial-state and final-state parton emission. The
parton ladder may be considered as a longitudinal color field or a flux tube, carrying transverse momentum of the hard scattering. The flux tubes expand and eventually get fragmented 
into string segments of quark-antiquark pairs. In case of many elementary parton-parton hard scattering in a collision, a large number of flux tubes are formed leading to a high local 
string-segment density, and subsequently high multiplicity of the collisional event. In the hydrodynamic EPOS3 model, the high local string-segment density, above a critical value, 
constitute the bulk matter or a medium. The string segments which do not have enough energy to escape from the bulk matter, form the ``core" that gets thermalized and undergoes 
(3 + 1D) viscus hydrodynamical evolution with a lattice QCD-complied cross-over equation of state and the ratio of the shear viscosity and entropy density, $\eta /s$ is taken as 0.08. 
The hydrodynamical evolution is followed by particle production in Cooper-Frye mechanism. After that, the hadronic evolution takes place and the ``soft" (low pT) hadrons freeze-out. 
The string segments from outside the bulk matter form the ``corona".  The string segments in the ``corona" hadronize by Schwinger's mechanism and escape as ``high" $p_{T}$ jet-hadrons. 
The string segments carrying enough energy to escape the bulk matter constitute the ``semi-hard" or intermediate $p_{T}$ particles. These segments, while escaping the bulk matter, 
pick-up quark or antiquark from within the bulk matter and the intermediate-$p_{T}$ hadrons thus produced in this process, inherit the properties of the bulk matter. After hadronization, 
the hadron-hadron re-scattering is modelled via UrQMD.\\

Using the EPOS 3.107 code, we have generated 40 million minimum-bias pp events for both the centre-of-mass energies, 7 and 13 TeV, for each of the options, with and without hydrodynamics. 
As presented in figure~\ref{fig:incl_spectra}, our simulated event sample from the hydrodynamic EPOS3 successfully describe the inclusive charged particle spectra from pp collisions at $\sqrt s$ 
= 7  \cite {ref28} and 13 TeV \cite {ref29}.  We analyze same sample of events in terms of the discussed observables. Suitable subsamples of different multiplicity classes and different kinematic 
cuts are selected from the simulated minimum-bias event samples. \\

\section{Analysis and Results}
\label{}
\subsection{Long-range ridge-like correlations}
\label{}
It will be pertinent to mention here, once again, that the ``ridge-like" two-particle long-range angular correlations, a signal of collective behaviour of particle production in heavy-ion collisions, 
as observed in $pp$ collision data at $\sqrt s$ = 7 TeV, could be qualitatively generated \cite {ref17} in ``flux-tube + hydro" approach, similar to the approach adopted in EPOS code. Inspite 
of the successes of the hydrodynamic approach in qualitative description, a quantitative comparison with the data is essential, particularly to access how the hydrodynamic description, 
implemented in EPOS, works better than the non-hydrodynamic models including the reasonably well understood particle production mechanism in the pQCD inspired multiple parton interaction 
 (MPI) model, like the one implemented in PYTHIA Monte Carlo Code.\\

The two-particle angular correlation function is defined by the per-trigger associated yields of charged particles obtained from $\Delta\eta,\Delta\varphi$ distribution (where $\Delta\eta$ 
and $\Delta\varphi$ are the differences in the pseudo-rapidity ($\eta$) and azimuthal angle ($\varphi$) of the two particles) and is given by:\\
\begin{equation}
\frac{1}{N_{trig}}\frac{d^{2} N^{assoc}}{d\Delta\eta d\Delta\varphi} = B(0,0)\times\frac{S(\Delta\eta, \Delta\varphi)}{B(\Delta\eta, \Delta\varphi)}
\label{Eq1}
\end{equation}
where $N_{trig}$ is the number of trigger particles in the specified $p_{T}^{trigger}$ range. \\

The function S($\Delta\eta, \Delta\varphi$) is the differential measure of per-trigger distribution of associated hadrons in the same-event, i.e,\\

\begin{equation}
S(\Delta\eta, \Delta\varphi) = \frac{1}{N_{trig}}\frac{d^{2} N^{assoc}_{same}}{d\Delta\eta d\Delta\varphi} \\
\end{equation}

The background distribution function B($\Delta\eta, \Delta\varphi$) is defined as: 
\begin{equation}
B(\Delta\eta, \Delta\varphi) =\frac{d^{2} N^{mixed}}{d\Delta\eta d\Delta\varphi}
\end{equation}
where $N^{mixed}$ is the number of mixed event pairs.\\

The factor B(0,0) in Eqn.~\ref{Eq1} is used to normalize the mixed-event correlation function such that it is unity at ($\Delta\eta, \Delta\varphi$)=(0,0). \\

\begin{figure}[htb!]
\centering
\includegraphics[scale=0.45]{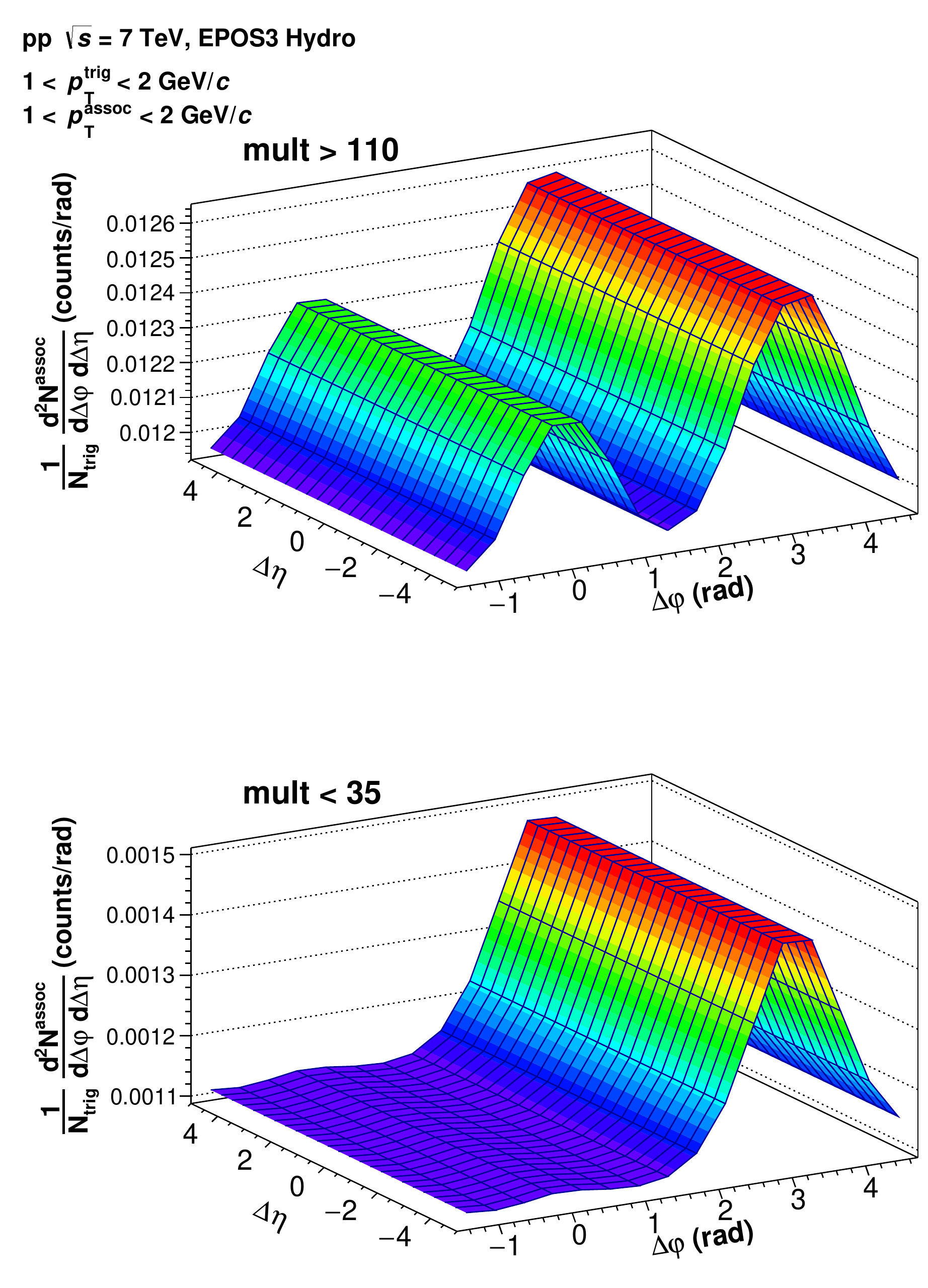}
\caption{Two particle ${\Delta\eta - \Delta\varphi}$ charge particle correlation function for 1 $\textless p_{T}^{trigger}, p_{T}^{associated} \textless$ 2 GeV/c with unidentified charged
particle as trigger, for the hydrodynamic-EPOS3 generated $pp$ collisions at $\sqrt s$ = 7 TeV for events of multiplicity-class $N_{ch} \textgreater 110$ (upper panel) and $N_{ch} 
\textless 35$ (lower panel). The short-range correlations have been suppressed for clear presentation of the long-range correlations} 
\label{fig:2D7TeV} 
\end{figure}

The two-particle azimuthal correlations study helps extracting several sources of correlations in multiparticle production, depending on the studied ranges of $|\Delta\eta|$ and also 
the $p_{T}$ for the trigger and the associated particles. In the context of the present study, the correlated emission of particles from collective medium can be extracted by studying 
the long-range ($|\Delta\eta | \gg 0$) two-particle azimuthal angle correlations. In relativistic heavy-ion collisions, the long-range two-particle azimuthal angle correlations are attributed 
to the formation of collective medium. The correlated pair yields per trigger with small $|\Delta\varphi |$ over a wide range of $|\Delta\eta |$ (long-range), result a ``ridge" structure in 
the constructed correlation functions. The analysis \cite {ref10} of LHC $pp$ data in terms of correlated yields as a function $|\Delta\varphi |$ reveals that the ``ridge"- structure becomes
prominent in pp collisions with increasing multiplicity of events. It is the near-side ($|\Delta\varphi | \sim 0$ ) long-range correlations, that are of particular interest for the present study of 
quantitative comparison of data and the hydrodynamic simulation of multi-particle production in high-multiplicity $pp$ events.\\

\begin{figure}[htb!]
\centering
\includegraphics[scale=0.55]{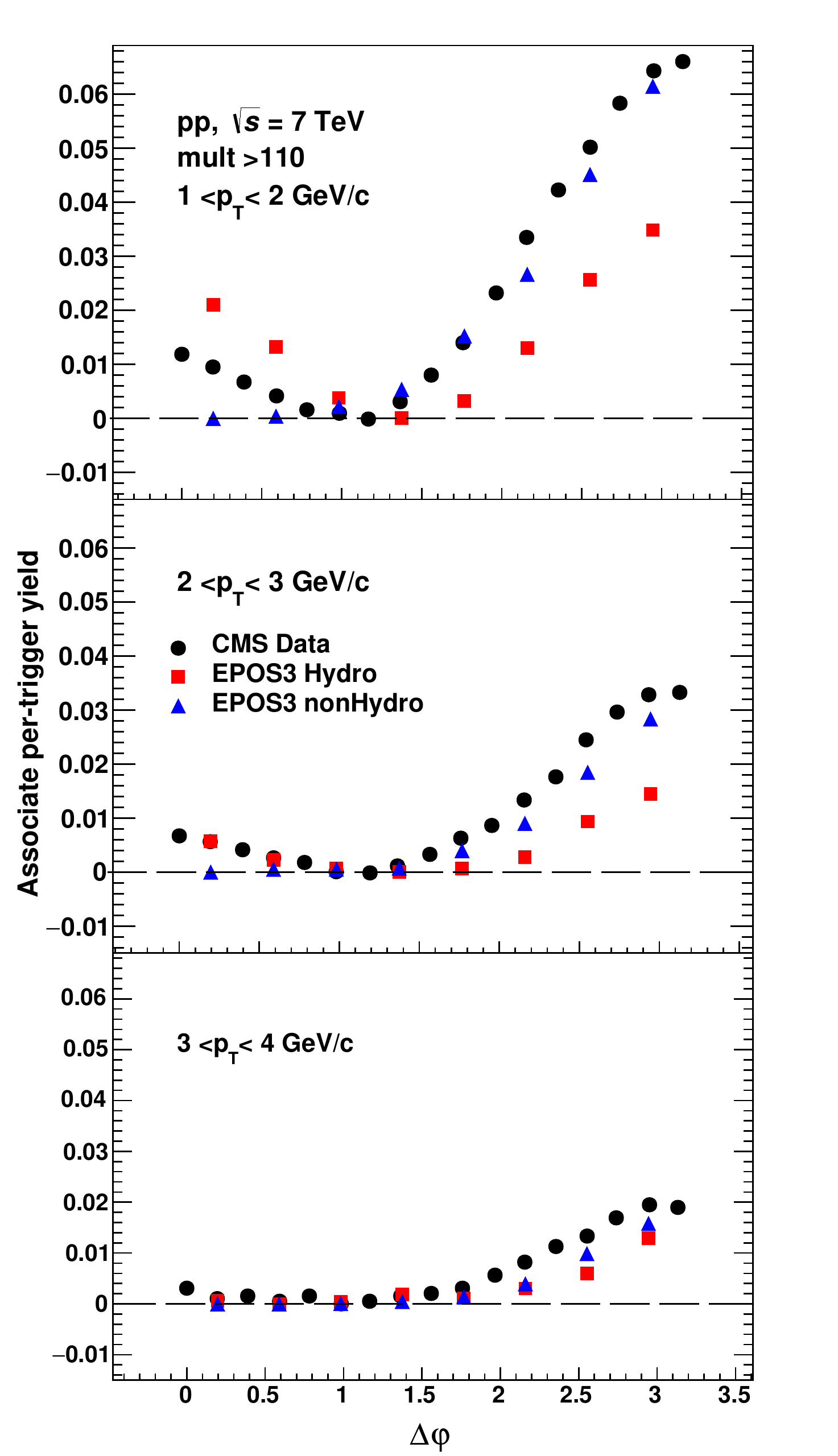}
\caption{One dimensional $\Delta\varphi$ projection for high-multiplicity events for the region of ridge-like correlations obtained from the long-range two-particle azimuthal correlations 
of charged particles, averaged over 2 $\textless |\Delta\eta| \textless$ 4, for 1 $\textless p_{T}^{trigger}, p_{T}^{associated} \textless$ 2 GeV/c, 2 $\textless p_{T}^{trigger}, p_{T}^{associated} 
\textless$ 3 GeV/c and 3 $\textless p_{T}^{trigger}, p_{T}^{associated} \textless$ 4 GeV/c from the data \cite {ref10} and the hydrodynamic-EPOS3 generated events of $pp$ collisions at 
$\sqrt s$ = 7 TeV.}
\label{fig:ridge7tev}
\end{figure}

\begin{figure}[htb!]
\centering
\includegraphics[scale=0.55]{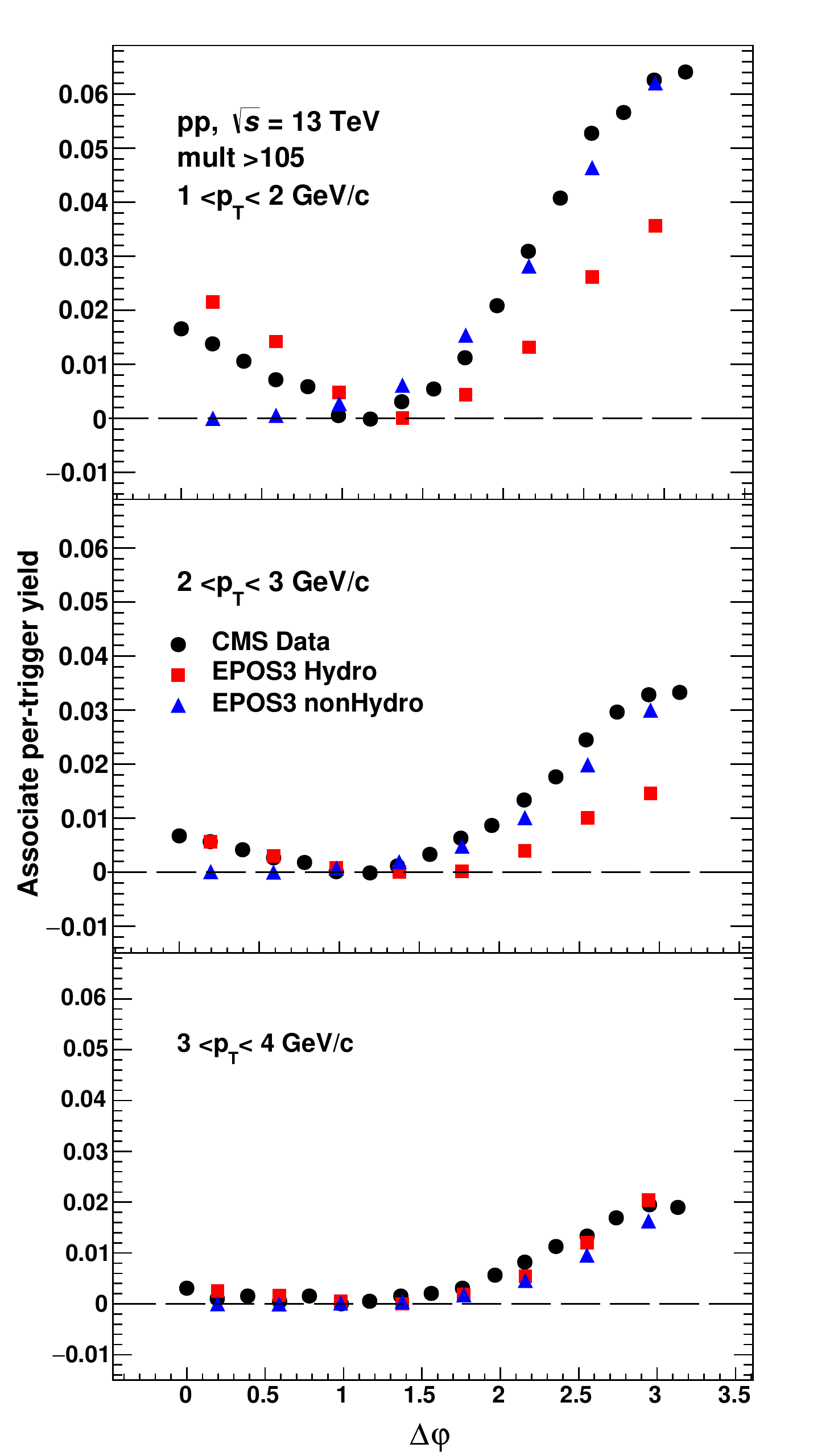}
\caption{Same as figure~\ref{fig:ridge7tev} for $pp$ events at $\sqrt s$ = 13 TeV.}
\label{fig:ridge13tev}
\end{figure}

We construct the long-range two-particle angular correlations of the charged particles in simulated events, matching the kinematic cuts and multiplicity classes as chosen for analysis
of the data at references - \cite {ref10} for $\sqrt s$ = 7 and 13 TeV. As expected from a hydrodynamic code of particle production like the EPOS3-hydro, the long-range two-particle 
azimuthal correlations of charged particles reveal a prominent ridge-like structure for high-multiplicity events, while such structure is absent in low-multiplicity events. Thc figure~\ref{fig:2D7TeV} 
contains representative plots of two-particle correlation function for 1 $\textless p_{T}^{trigger}, p_{T}^{associated} \textless$ 2 GeV/c with unidentified charged particle as trigger, for the 
hydrodynamic-EPOS3 generated $pp$ collisions at $\sqrt s$ = 7 TeV for events of multiplicity-classes $N_{ch} \textgreater 110$ and $N_{ch} \textless 35$, after removing the short-range 
jet-like correlations. The per-trigger correlated yield, for representative high-multiplicity event classes, in different $p_{T}$-intervals, for both the non-hydrodynamic and hydrodynamic EPOS3 
generated $pp$ collisions at $\sqrt s$ = 7 and 13TeV are projected onto $\Delta\varphi$, after subtracting the ``zero yield at minimum", $Yield_{|ZYAM|}$ \cite {ref10}, and are 
shown in the  figure~\ref{fig:ridge7tev} and figure~\ref{fig:ridge13tev}, respectively, to compare the data in the same kinematic ranges. \\

The appearance of the ridge-like structure in the long-range two-particle angular correlations of the charged particles in the high-multiplicity EPOS3-hydro generated $pp$ events 
at $\sqrt s$ = 7 and 13 TeV reflects the collective property, that is expected in a hydrodynamic model of particle production. The correlated yields of high-multiplicity event class 
as a function of $\Delta\varphi$ for different $p_{T}$-intervals in the simulated events reveals similar feature as observed in the two-particle azimuthal correlations of the charged 
particles in the data: the ridge-like structure is most prominent in the 1 to 2 GeV/c $p_{T}$-range and in the highest-multiplicity events, while it gradually decreases with increasing 
$p_{T}$. Nevertheless, as it is clear in the  figure~\ref{fig:ridge7tev} and  figure~\ref{fig:ridge13tev}, for the most prominent $p_{T}$-range of 1-2 GeV/c, the EPOS3 events 
overestimate the correlated yields as compared to the data. \\

\subsection {Blast wave parametrization}
\label{}
In a hydrodynamic picture of relativistic collisions of heavy nuclei, the collective radial flow, generated due to the pressure gradient in the system, is reflected in the spectra of identified 
final state charged particles produced in relativistic collisions. The Boltzmann-Gibbs blast-wave (BGBW) \cite {ref21} model, a hydrodynamics-motivated empirical formalism, estimates 
the radial flow by analyzing the identified particle spectra. The blast-wave model considers that the particles produced in the collision are locally thermalized and the system expands 
collectively with a common velocity field. Though model does not include hydrodynamic evolution, it considers that the system undergoes an instantaneous common freeze-out at a kinetic 
freeze-out temperature ($T_{kin}$) and a transverse radial flow velocity ($\beta$) at the freeze-out surface. The BGBW, thoroughly used in analyzing the relativistic nucleus-nucleus collisions 
data, revealed \cite {ref19} transverse radial flow for high-multiplicity $pp$ collisions \cite {ref20} data, also. \\

Assuming the hard-sphere particle source of uniform density, the transverse momentum spectra, in the BGBW model, is given by, 
\begin{equation}
\frac{dN}{p_{T}dp_{T}} \propto \int_{0}^{R} rdr \ m_{T} \ \bold I_{o} \left(\frac{p_{T}Sinh \ \rho}{T_{kin}}\right)  \bold K_{1}\left(\frac{m_{T}Cosh \ \rho}{T_{kin}}\right)
\end{equation}  
where $\rho$ = $ tan h^{-1}\beta$, $\bold I_{0}$ and $\bold K_{1}$ are modified Bessel functions. \\

The flow velocity profile is given by,
\begin{equation}
\beta = \beta_{s}(\frac{r}{R})^{n}
\end{equation}  
where $\beta_{s}$ is the surface velocity and $r/R$ is the relative radial position in the thermal source. The average transverse flow velocity,
$\langle\beta\rangle$ is given by, $\langle\beta\rangle  =  \frac{2}{(2+n)}$$\beta_{s}$.\\

\begin{figure}[htb!]
\centering
\includegraphics[scale=0.45]{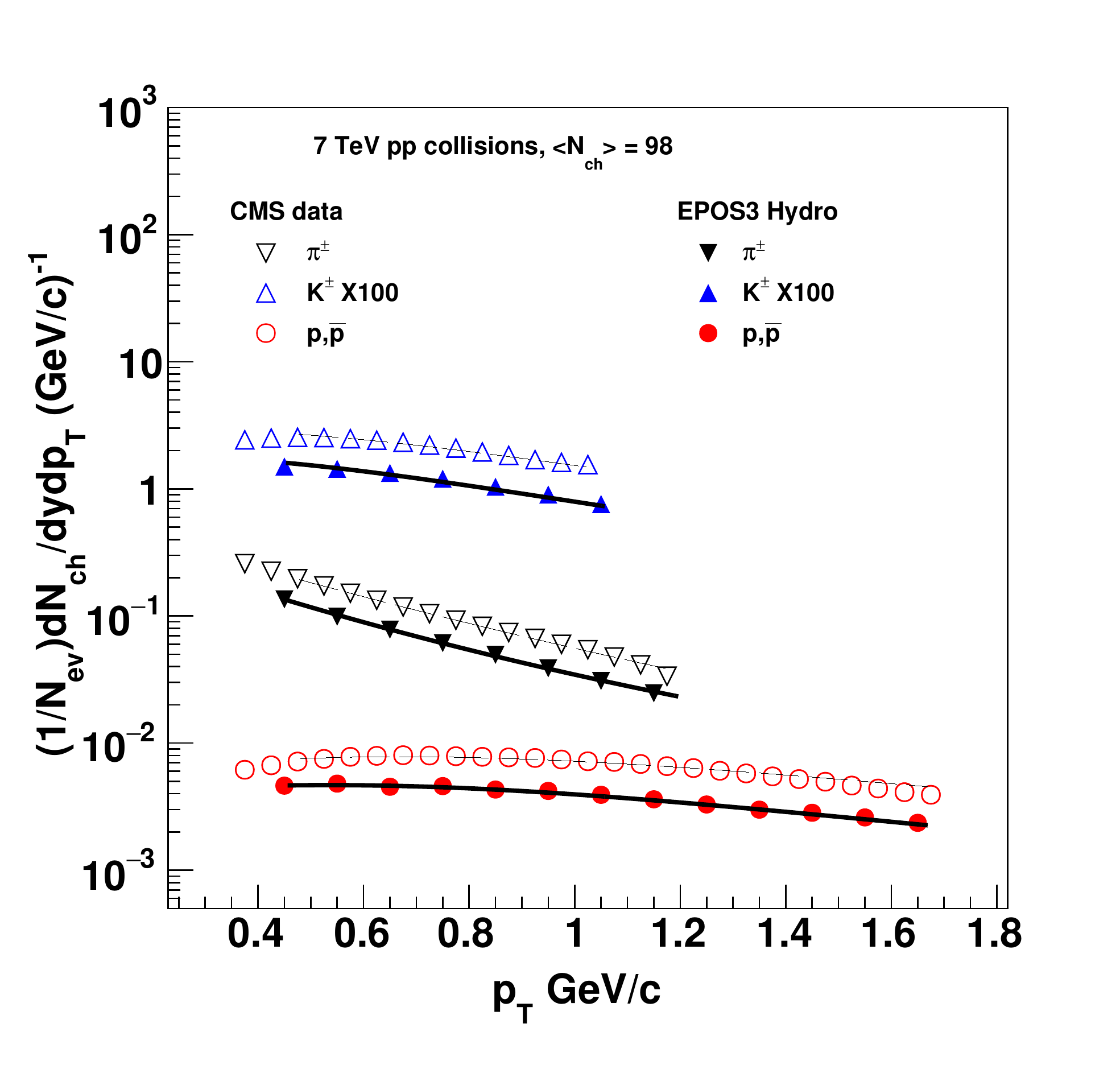}
\caption{The transverse momentum spectra for $\pi^\pm$, $K^\pm$, $p$($\bar p$) as measured by the CMS experiment \cite{ref20} at LHC for the event-class of average 
multiplicity = 98 in $pp$-collisions at $\sqrt {s}$ = 7 TeV, along with BG-blast-wave fits (solid lines). The uncorrelated statistical and systematic uncertainties have been added in quadrature.}
\label{fig:BW_7_98} 
\end{figure}

The hydrodynamic EPOS3 generated $pp$ events are expected to exhibit the radial flow. Using the BGBW formalism, we intend to compare quantitatively  the radial flow 
parameters for the EPOS3 generated events and the data. We use the Chi-square ($\chi^2$) test to ensure goodness of fit while obtaining the fit-parameters, 
the kinetic freeze-out temperature and the radial flow velocity, from the spectra data as well as from the generated spectra. For our analysis, we put lower $p_{T}$ - cut at 
0.475 GeV/c of spectra for all the species. At the higher side, the $p_{T}$-range is limited to $p_{T} < 2$ GeV/c or less, depending on the availability of the data. \\

We calculate $R$ for different event classes with different $\langle N_{ch} \rangle$ in $|\eta| <2.4$ in this study, from the relation, $R(\langle N_{ch} \rangle) = a. \langle N_{ch} 
\rangle^{1/3}$ where $a = 0.597 \pm 0.009 (stat.) \pm 0.057(syst.)$ fm at 0.9 TeV and $a = 0.612 \pm 0.007 (stat.) \pm 0.068(syst.)$ fm, as parameterized \cite {ref20} by the 
CMS experiment from the measurement of radius of source of emission as a function of average charged particle multiplicity for $pp$ collisions at $\sqrt s$ = 7 TeV. The BGBW 
fit parameters are available \cite {ref19} for different multiplicity classes of pp collisions at  $\sqrt s$ = 7 TeV \cite {ref20}. We fit the blast-wave function to the $p_{T}$ - spectra 
for different sets of data for $\sqrt s$ = 13 TeV \cite {ref30}, keeping the kinetic freeze-out temperature ($T_{kin}$), the radial flow velocity ($\beta_{s}$) and the exponent 
($n$) of the flow velocity profile free to produce the best possible simultaneous or combined fits to the data, in terms of $\chi^2$/ndf.\\

\begin{figure}[htb!]
\centering
\includegraphics[scale=0.45]{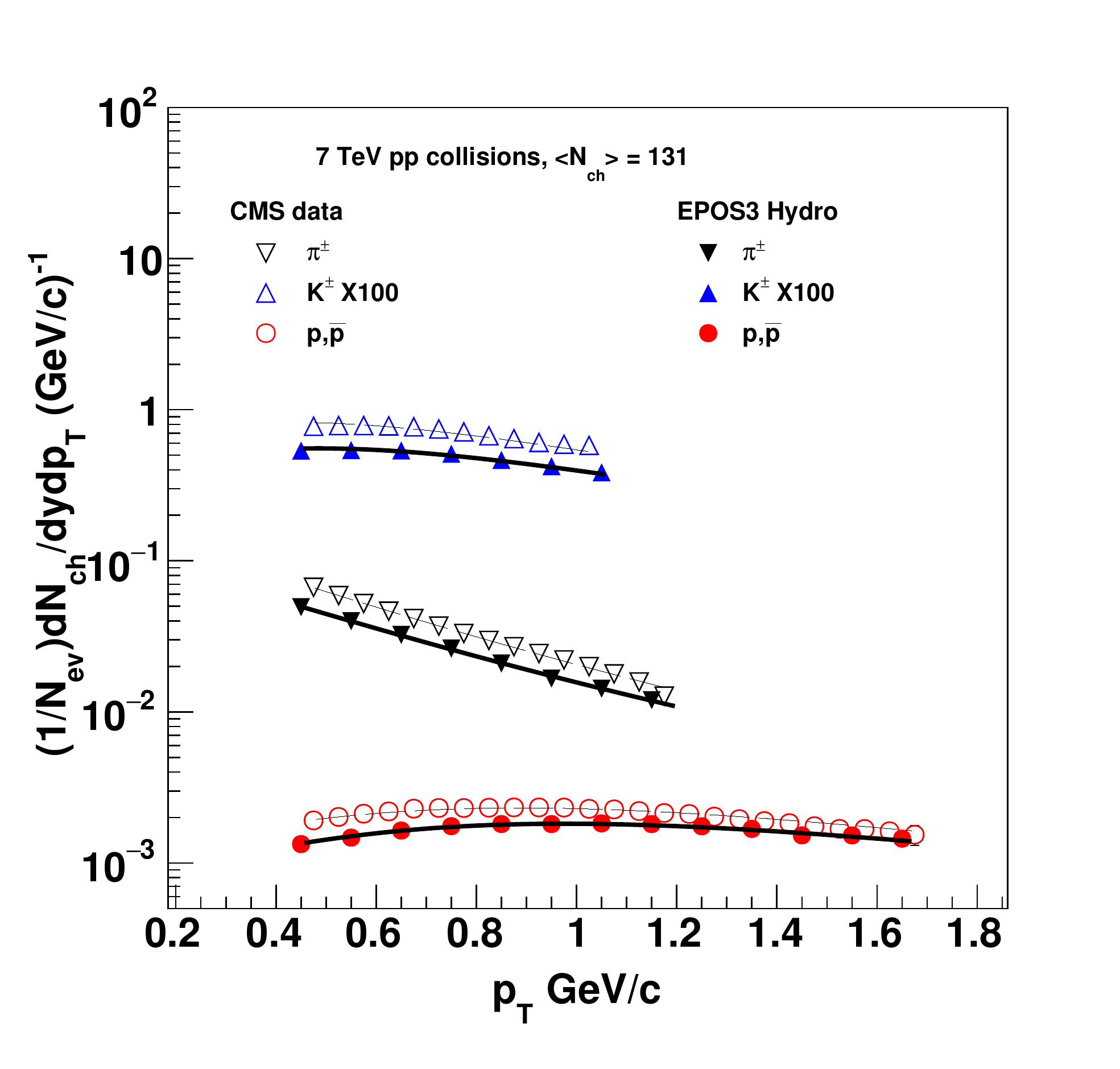}
\caption{The transverse momentum spectra for $\pi^\pm$, $K^\pm$, $p$($\bar p$) as measured by the CMS experiment \cite {ref20} at LHC for the event-class of average multiplicity 
= 131 in $pp$-collisions at $\sqrt {s}$ = 7 TeV, along with BG-blast-wave fits (solid lines). The uncorrelated statistical and systematic uncertainties have been added in quadrature.}
\label{fig:BW_7_131} 
\end{figure}

The figures~\ref{fig:BW_7_98} and ~\ref{fig:BW_7_131} are representative plots where BGBW fits to the spectra data of identified particles from pp collisions at 7 TeV for the event-classes 
of average multiplicity = 98 and 131, respectively, are shown along with the fits to the simulated spectra, obtained from the respective class of simulated events. \\

\begin{table}[h]
\begin{center}
\label{tab}
\scalebox{0.8}{
\begin{tabular}{|c|c|c|c|c|c|}
\hline
$\sqrt {s} (TeV)$&$\langle N_{ch} \rangle$& $T_{kin} (MeV) $& $\langle \beta \rangle$& $n$&$\chi^2/n.d.f$ \\
\hline
$ 7$ &$98$&$115.57\pm 0.11$ & $ 0.766\pm 0.004$ & $ 0.540\pm 0.006$&$1.02$\\
\hline
$ 7$ &$109$&$113.09\pm 0.12$ & $ 0.779\pm 0.004$ & $ 0.503\pm 0.006$&$0.61$\\
\hline
$ 7$ &$120$&$110.84\pm 0.15$ & $ 0.790\pm 0.004$ & $ 0.480\pm 0.006$&$0.34$\\
\hline
$ 7$ &$131$&$104.29\pm 0.15$ & $ 0.809\pm 0.005$ & $ 0.436\pm 0.005$&$0.44$\\
\hline
$ 13$ &$108$&$140.80\pm 0.022$ & $ 0.723\pm 0.005$ & $ 0.58\pm 0.01$&$1.65$\\
\hline
$ 13$ &$119$&$129.31\pm 0.019$ & $ 0.778\pm 0.002$ & $ 0.56\pm 0.011$&$0.86$\\
\hline
$ 13$ &$130$&$128.29\pm 0.019$ & $ 0.763\pm 0.004$ & $ 0.50\pm 0.009$&$1.18$\\
\hline
$ 13$ &$141$&$119.77\pm 0.016$ & $ 0.764\pm 0.004$ & $ 0.48\pm 0.01$&$1.40$\\
\hline
$ 13$ &$151$&$112.84\pm 0.016$ & $ 0.783\pm 0.004$ & $ 0.44\pm 0.011$&$1.44$\\
\hline
$ 13$ &$162$&$102.67\pm 0.017$ & $ 0.826\pm 0.003$ & $ 0.36\pm 0.007$&$0.93$\\
\hline
\end{tabular}
}
\caption{$T_{kin}$, $\langle\beta\rangle$ and $n$, the parameters of the the BGBW, obtained from the simultaneous fit to the published \cite{ref20, ref29} spectra of  $\pi^\pm$, 
$K^\pm$ and $p$($\bar p$) and respective $\chi^2 /n.d.f$ for $pp$ collisions at $\sqrt {s}$ = 7 and 13 TeV for different event classes depending on average multiplicity, 
$\langle N_{ch} \rangle$, in the range $|\eta| < 2.4$.}
\label{tab:data_BW}
\end{center}
\end{table}

We tabulate the fit parameters, the kinetic freeze-out temperature ($T_{kin}$), the average radial flow velocity  ($\langle\beta\rangle$) at the freeze-out surface, and the exponent ($n$) 
as obtained by simultaneous fit of identified particle spectra by BGBW for different classes of high-multiplicity pp events and for $\sqrt {s}$ = 7 and 13 TeV, along with respective $\chi^2/ n.d.f$ 
in table~\ref{tab:data_BW}. The table includes parameters for those event classes which fit reasonably with the BW function. The table~\ref{tab:epos_BW} presents similar results for the 
EPOS3-hydro model generated events.\\

The figures~\ref{fig:BW_7_98} and ~\ref{fig:BW_7_131} along with the tables~\ref{tab:data_BW} and ~\ref{tab:epos_BW} show that the data and the EPOS3-hydro generated events are far from
quantitative agreement, in terms of the BW-parameters, except for very-high multiplicity classes of average multiplicity = 151 and 162 in $pp$-collisions at $\sqrt {s}$ = 13 TeV.\\

\begin{table}[h]
\begin{center}
\label{tab}
\scalebox{0.8}{
\begin{tabular}{|c|c|c|c|c|c|}
\hline
$\sqrt {s} (TeV)$&$\langle N_{ch} \rangle$& $T_{kin} (MeV) $& $\langle \beta \rangle$& $n$&$\chi^2/n.d.f$ \\
\hline
$ 7$ &$98$&$106.10\pm 0.015$ & $ 0.768\pm 0.0003$ & $ 0.59\pm 0.001$&$30.83$\\
\hline
$ 7$ &$109$&$105.91\pm 0.008$ & $ 0.808\pm 0.001$ & $ 0.46\pm 0.005$&$1.51$\\
\hline
$ 7$ &$120$&$103.30\pm 0.01$ & $ 0.813\pm 0.002$ & $ 0.45\pm 0.003$&$1.59$\\
\hline
$ 7$ &$131$&$103.02\pm 0.02$ & $ 0.829\pm 0.002$ & $ 0.39\pm 0.004$&$0.52$\\
\hline
$ 13$ &$108$&$142.00\pm 0.002$ & $ 0.749\pm 0.001$ & $ 0.63\pm 0.01$&$7.54$\\
\hline
$ 13$ &$119$&$142.00\pm 0.0019$ & $ 0.774\pm 0.005$ & $ 0.50\pm 0.002$&$0.86$\\
\hline
$ 13$ &$130$&$141.96\pm 0.006$ & $ 0.774\pm 0.008$ & $ 0.45\pm 0.009$&$1.16$\\
\hline
$ 13$ &$141$&$127.98\pm 0.016$ & $ 0.797\pm 0.001$ & $ 0.44\pm 0.004$&$1.31$\\
\hline
$ 13$ &$151$&$112.90\pm 0.01$ & $ 0.814\pm 0.006$ & $ 0.43\pm 0.004$&$0.96$\\
\hline
$ 13$ &$162$&$100.52\pm 0.015$ & $ 0.815\pm 0.007$ & $ 0.42\pm 0.007$&$1.43$\\
\hline
\end{tabular}
}
\caption{$T_{kin}$, $\langle\beta\rangle$ and $n$, the parameters of the the BGBW, obtained from the simultaneous fit to the spectra obtained from simulated EPOS events 
for $\pi^\pm$, $K^\pm$ and $p$($\bar p$) and respective $\chi^2 /n.d.f$ for $pp$ collisions at $\sqrt {s}$ = 7 and 13 TeV for different event classes depending on average 
multiplicity, $\langle N_{ch} \rangle$, in the range $|\eta| < 2.4$.}
\label{tab:epos_BW}
\end{center}
\end{table}

\subsection{The $\langle p_{T} \rangle$ as a function of $\langle N_{ch} \rangle$}
\begin{center}
\begin{figure}
\includegraphics[scale=0.45]{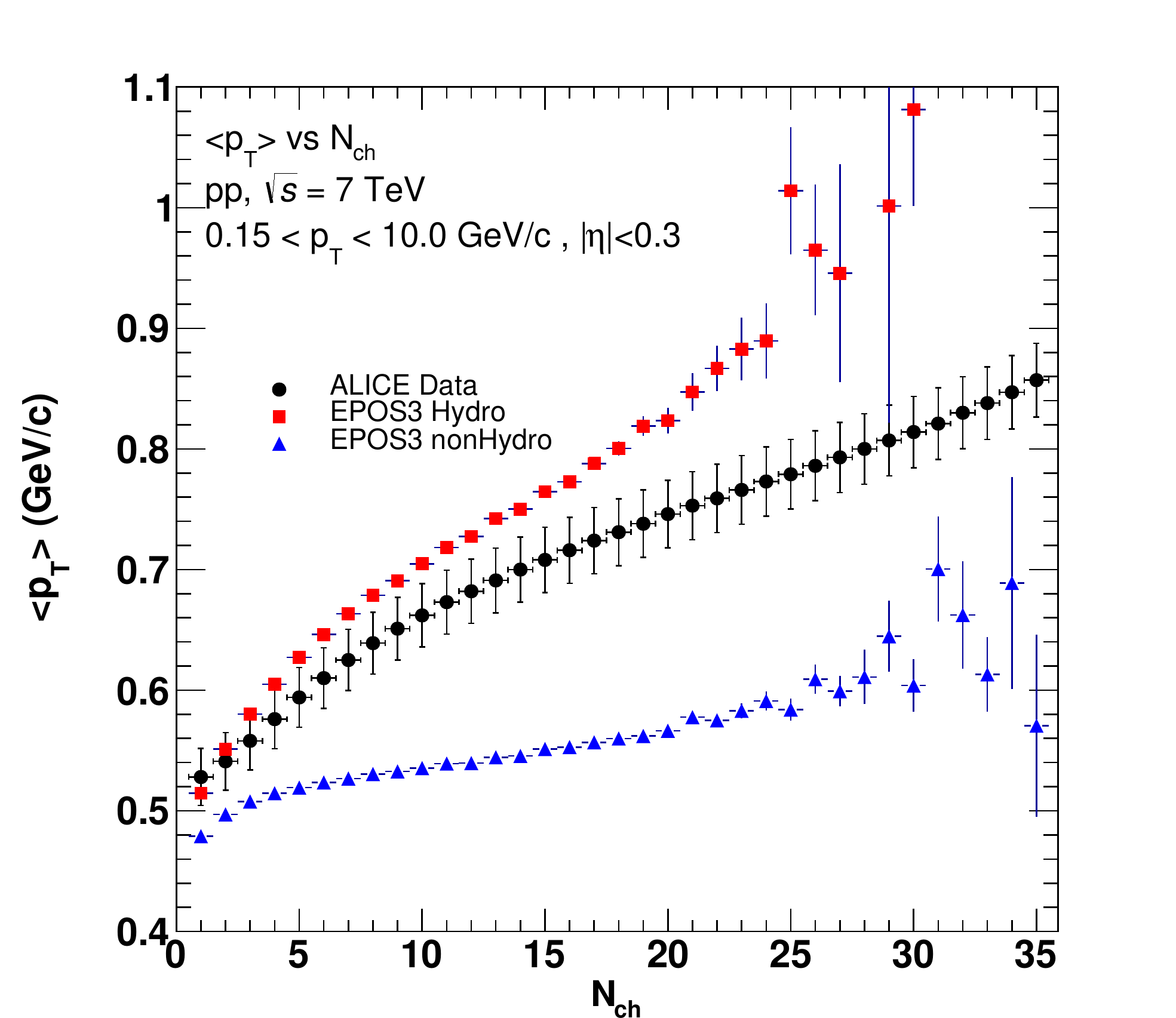}
\caption{Average transverse momentum, $\langle p_{T} \rangle$, as a function of charged particle multiplicity, $N_{ch}$, as measured \cite {ref31} by ALICE is compared with 
the simulated events from EPOS3 event generator, with and without Hydro calculations.}
\label{fig:alice_setup} 
\end{figure}
\end{center}

The ALICE experiment at the LHC measured \cite {ref31} on $\langle p_{T} \rangle$ of charged particles, in the pseudorapidity range $|\eta| <1.0$ and with the transverse momentum, 
$p_{T}$ up to 10 GeV/c, as a function of $\langle N_{ch} \rangle$ and showed that data of $pp$ collisions at $\sqrt {s}$ = 7 TeV could be well described by the pQCD-inspired multiple 
parton interaction (MPI) model with color reconnection, as implemented in PYTHIA monte carlo code. We calculate the observables, in the the same kinematic ranges used by the ALICE, 
for the events generated with both the hydro and non-hydro EPOS3 simulations. The results are depicted in the plots, along with the data in Fig.~\ref{fig:alice_setup}. \\

\begin{center}
\begin{figure}
\includegraphics[scale=0.45]{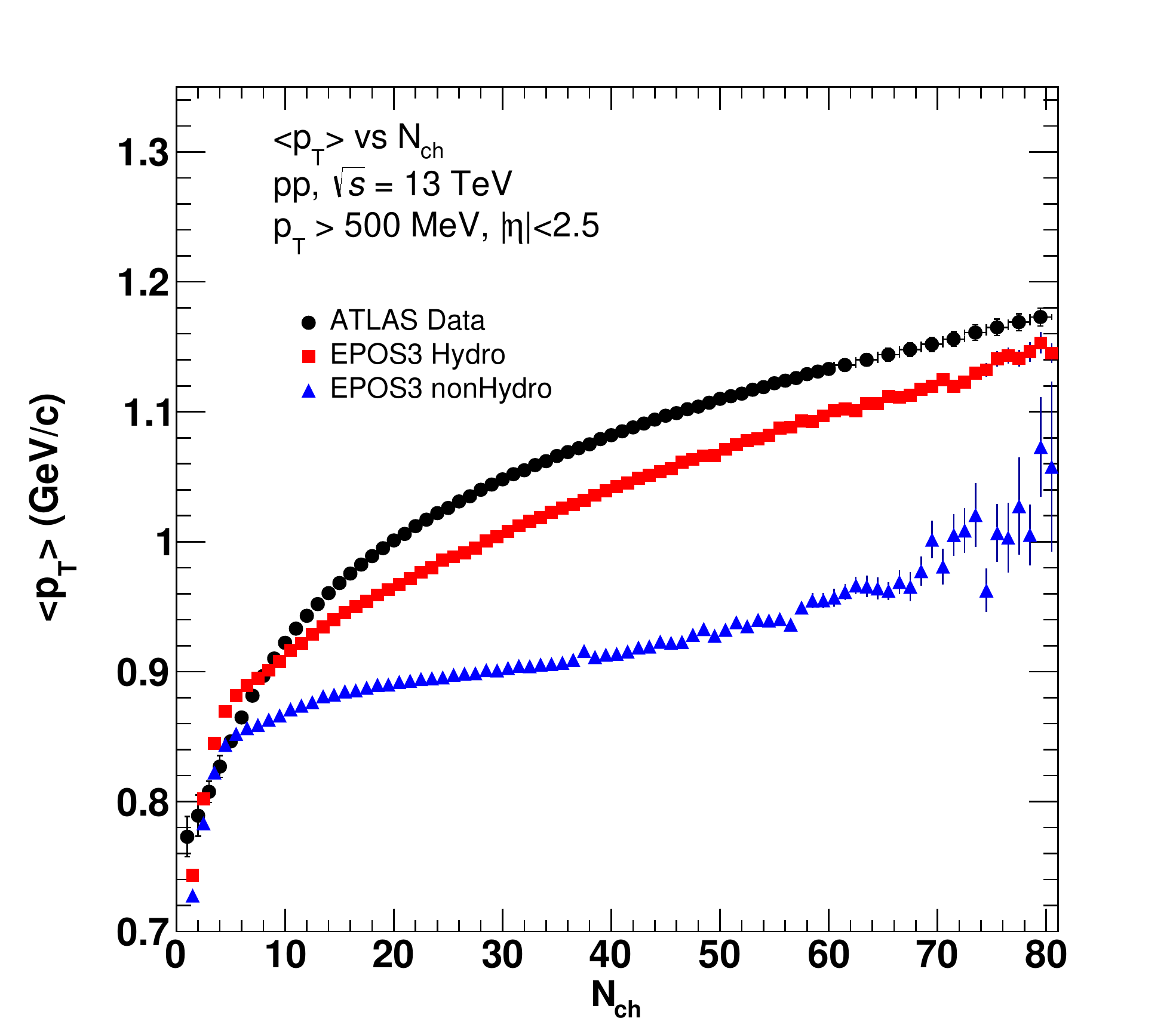}
\caption{Average transverse momentum, $\langle p_{T} \rangle$, as a function of charged particle multiplicity, $N_{ch}$, as measured \cite {ref28} by ATLAS is compared with 
the simulated events from EPOS3 event generator, with and without Hydro calculations.}
\label{fig:atlas_setup} 
\end{figure}
\end{center}

As is clear from Fig.~\ref{fig:alice_setup}, the EPOS3 code, with or without hydrodynamics, cannot describe the ALICE measurement of $\langle p_{T} \rangle$ as a function of 
$\langle N_{ch} \rangle$. The ATLAS experiment has studied \cite {ref28} the same for $pp$ collisions at $\sqrt {s}$ = 13 TeV, though in different kinematic ranges, $|\eta| <2.4$ and $p_{T} > 0.5$ 
GeV/c. We repeat the $\langle p_{T} \rangle$ as a function of $\langle N_{ch} \rangle$ analysis in accordance with the kinematic cuts used for the ATLAS data. The comparison of the 
simulation and the data is presented in Fig.~\ref{fig:atlas_setup}. In this case also the data and the simulated events vary widely. It may be noted that both the ALICE and the ATLAS data 
includes particles of $p_{T}$, much higher than the $p_{T}$-range of the``soft" particles likely to be produced from the ``core" or the bulk collective medium that is considered in the EPOS 
hydrodynamic code. In view of this, to compare the data with ``soft" particles only, we choose the CMS data on identified $p_{T}$-spectra from events of different multiplicity classes.  \\

The CMS experiment has measured $p_{T}$-spectra of $\pi^\pm$, $K^\pm$, $p$($\bar p$) over the rapidity, ($y = (1/2) ln\frac{E + p_{L}}{E - p_{L}}$) range $|y|<1$ for the $pp$ collisions at 
$\sqrt {s}$ = 7 and 13 TeV $|\eta| <2.4$. The measured $p_{T}$ - ranges for the measured identified particles in the $pp$ collisions at both the energies, 7 TeV \cite {ref20} and 13 TeV \cite {ref30}
are 0.1 to 1.2 GeV/c for $\pi^\pm$, 0.2 to 1.050 GeV/c for $K^\pm$ and 0.35 - 1.7 GeV/c for $p$ and $\bar p$. The measured $p_{T}$-ranges fall within the $p_{T}$ - range of EPOS3 for particles 
originating from hydrodynamic bulk medium.\\

\begin{center}
\begin{figure}
\includegraphics[scale=0.55]{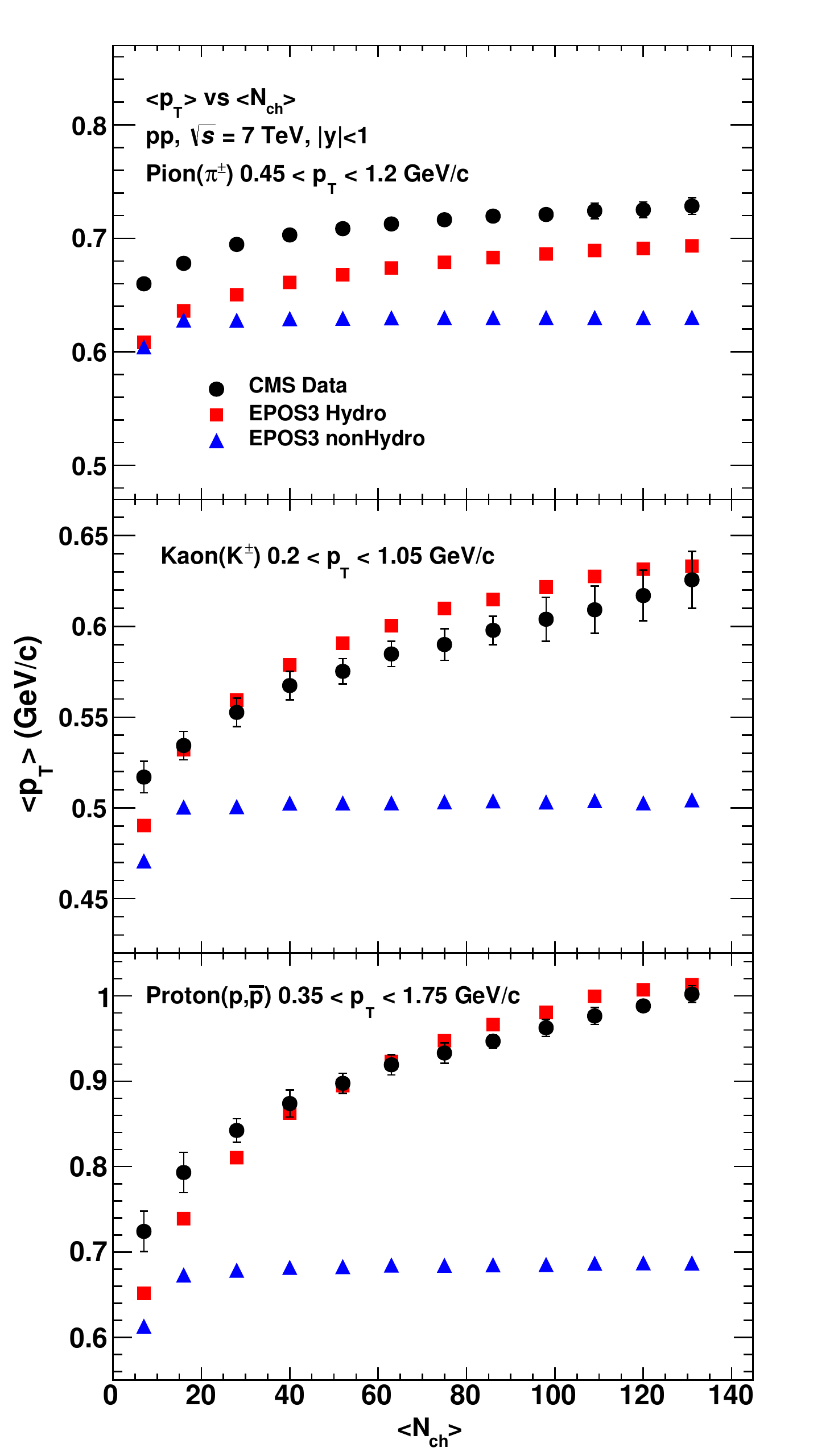}
\caption{Average transverse momentum, $\langle p_{T} \rangle$, as a function of mean charged particle multiplicity, $\langle N_{ch} \rangle$, for the identified 
charged particles in $pp$ collisions at $\sqrt {s} =$ 7 TeV. The CMS data \cite {ref20}, have been compared with simulated events using EPOS3 event 
generator with and without hydrodynamics.}
\label{fig:meanptVsNch_piKp_7TeV} 
\end{figure}
\end{center}

\begin{center}
\begin{figure}
\includegraphics[scale=0.55]{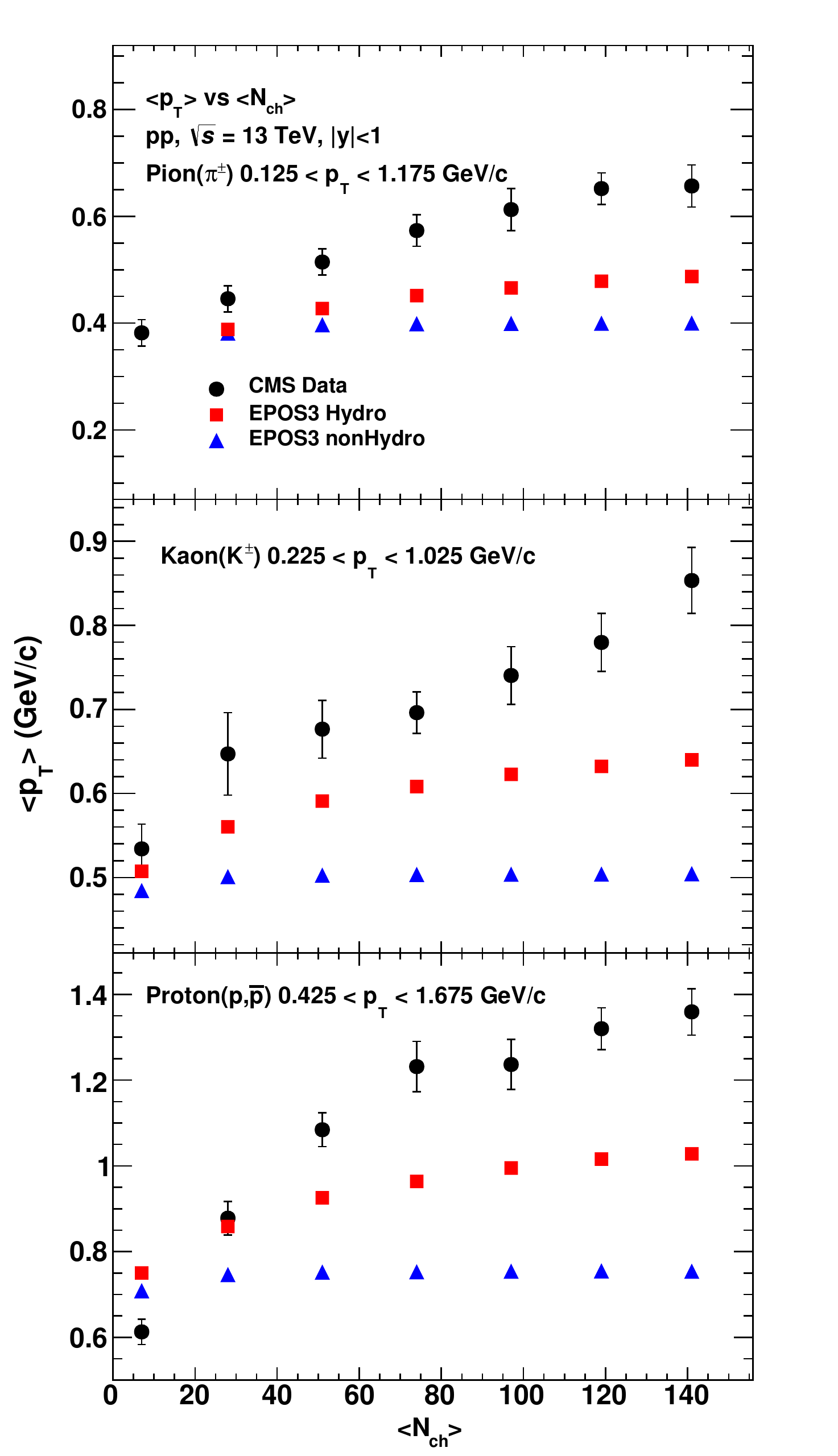}
\caption{Average transverse momentum, $\langle p_{T} \rangle$, as a function of mean charged particle multiplicity, $\langle N_{ch} \rangle$, for the identified 
charged particles in $pp$ collisions at $\sqrt {s} =$ 13 TeV. The CMS data \cite {ref30}, have been compared with simulated events using EPOS3 event 
generator with and without hydrodynamics.}
\label{fig:meanptVsNch_piKp_13TeV} 
\end{figure}
\end{center}

We compute $\langle p_{T} \rangle$ from the CMS data on identified charged particle spectra for different event classes from $pp$ collisions at $\sqrt {s}$ = 7 \cite{ref20} and 
13 \cite {ref30} TeV. It is clear from figure.~\ref{fig:meanptVsNch_piKp_7TeV} that for pp collisions data at 7 TeV, the $\langle p_{T} \rangle$ of majority of produced ``soft" 
particles, the pions, for the simulated events do not match the measured ones in the given kinematic range. For the pp collisions at 13 TeV, the mismatch between the data
and the simulated events, as shown in figure~\ref{fig:meanptVsNch_piKp_13TeV}, is quite wide for all the identified particles.  \\

\subsection{Multiplicity dependent inverse slope parameter of $m_{T}$-distributions}
\label{}
The slope of transverse mass $m_{T}$-spectra of identified particle contains information of temperature of a medium, if formed, from where the particles are produced,
and the effect of transverse expansion of the medium. We obtain the $m_{T}$ spectra of identified charged particle of mass $m$ for different high-multiplicity classes of 
pp events at $\sqrt {s} =$ 7 and 13 TeV from the $p_{T}$-spectra measured \cite {ref20, ref30} by the CMS experiments from the relation, $m_{T} = (m^2 + p_{T}^2)^{1/2}$.

The $m_{T}$-spectra are fitted, in the range corresponding to low-$p_{T}$, with the exponential function:

\begin{equation}
\frac{dN}{m_{T}dm_{T}} = C.exp(- \frac{m_{T}}{T_{effective}})
\end{equation} 

where $T_{effective}$, the inverse slope parameter, contains the effect due to the transverse expansion of the system. The increase in the inverse slope parameter, $T_{effective}$ 
for the most commonly measured identified particles ($\pi^\pm$ , $K^\pm$, $p$ and $\bar p$), as has been observed in heavy-ion collisions \cite{ref32, ref33} is attributed to the 
collective transverse flow of the medium formed in the collision. The increase in inverse slope parameters has also been observed in high-multiplicity event classes of $pPb$ 
collisions at $\sqrt s_{NN} =$ 5.02 TeV \cite{ref34} and pp collisions at $\sqrt {s} =$ 7 \cite {ref35}.\\

\begin{center}
\begin{figure}
\includegraphics[scale=0.45]{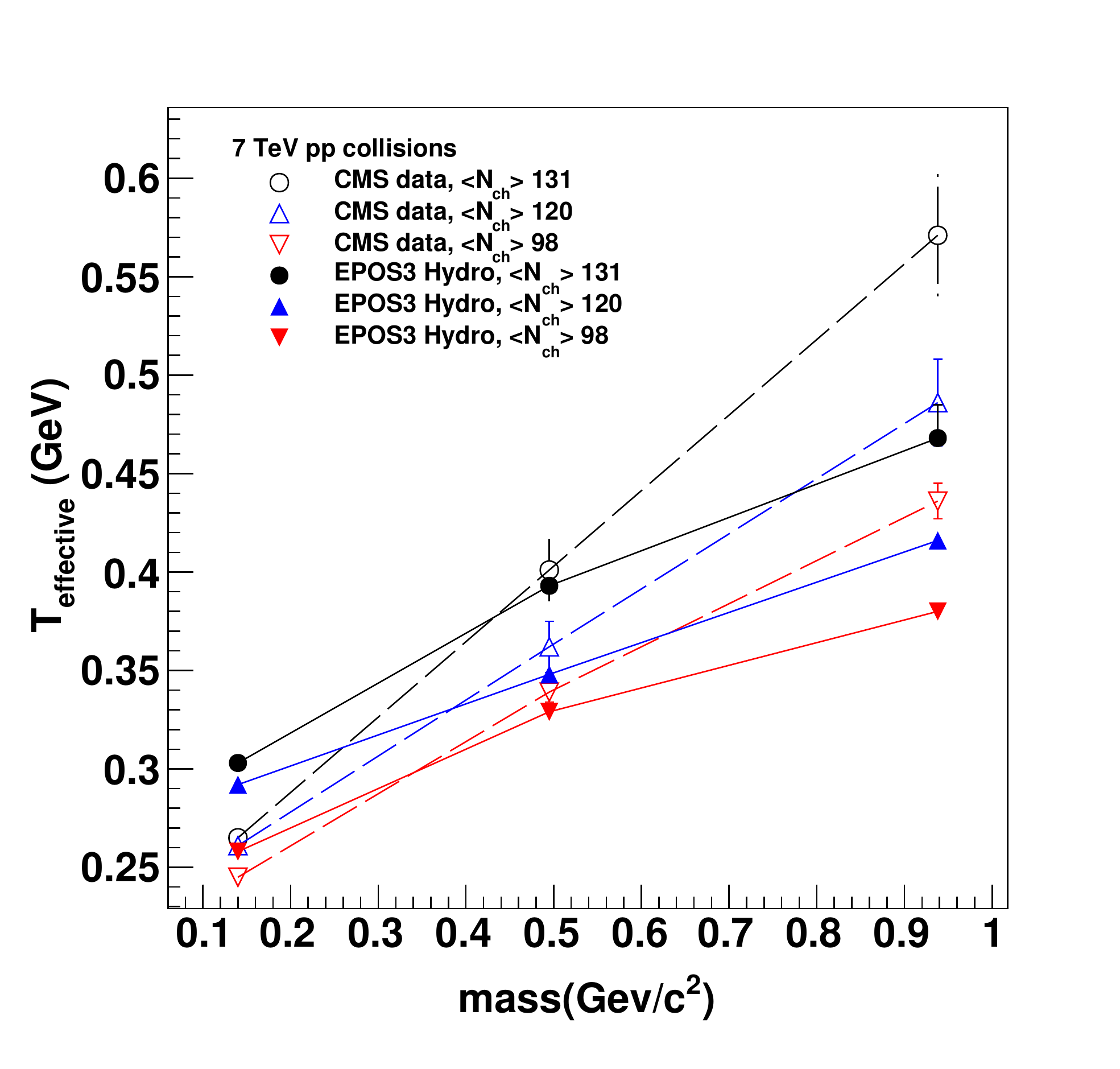}
\caption{The mass ordering of the inverse slope parameter $T_{effective}$ of identified particles ($m_{\pi^\pm} = 0.14$, $m_{K^\pm} = 0.495$, $m_{p(\bar p)} = 0.938$ $GeV/c^{2}$) as 
measured by the CMS experiment \cite {ref20} at $\sqrt {s}$ = 7 TeV, for a few event classes of high-multiplicity, is compared with those obtained from EPOS3-hydro simulated events. 
The $\langle N_{ch} \rangle$ is the mean multiplicity of the charged particles of respective event-class.}
\label{fig:massordering7} 
\end{figure}
\end{center}

\begin{center}
\begin{figure}
\includegraphics[scale=0.45]{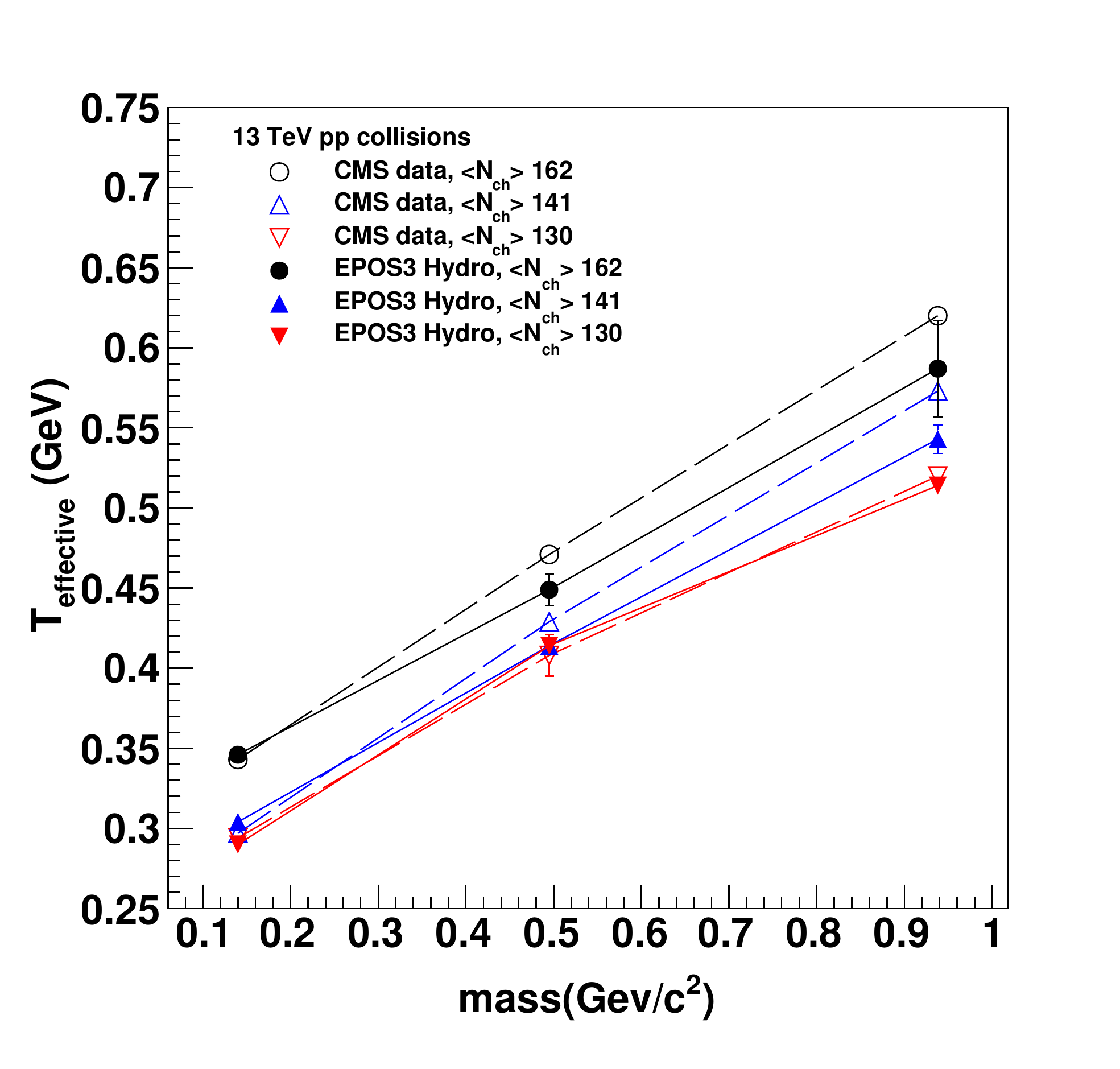}
\caption{Same as figure~\ref{fig:massordering7} for measured CMS data \cite {ref30} and EPOS3 simulated events of pp collisions at $\sqrt {s}$ = 13 TeV}
\label{fig:massordering13} 
\end{figure}
\end{center}

We fit the $m_{T}$ - spectra of identified particles in the overlapped range (0.475 $\textless p_{T} \textless$ 1.025) of $p_{T}$ - spectra at $\sqrt {s}$ = 7 and 13 TeV, obtained from 
the data \cite {ref20, ref30} as well as from the EPOS3-hydro simulation for the high-multiplicity event classes. The inverse slope parameters, obtained from the best fit of the spectra,
in terms of $\chi^2$/ndf, using the MINUTE program in the ROOT analysis framework \cite {ref29} are presented in figures~\ref{fig:massordering7} and ~\ref{fig:massordering13} for 
some representative high-multiplicity event-classes\\
 
From the figures~\ref{fig:massordering7} and ~\ref{fig:massordering13}, it is clear that while the EPOS3-hydro high-multiplicity $pp$ events exhibit mass ordering of inverse slope 
parameter of the $m_{T}$ - distributions, they largely deviate from the ones obtained from the measured spectra.\\

\section{Summary and Conclusion}
\label{}
In the context of several experimental signatures of collective nature of particle productions in relativistic collisions of small systems at RHIC and LHC, we analyse simulated
pp events at $\sqrt s$ = 7 and 13 TeV from EPOS3-hydro code, a hydrodynamic model of particle production in different systems of relativistic collisions, for quantitative 
comparisons with the particle production data in high-multiplicity $pp$ events for better understanding of the experimental signals.\\
 
We have studied the multiplicity dependence of the long-range two-particle angular correlations of charged particles, produced in EPOS3-generated $pp$ events. The EPOS3-hydro 
generated high-multiplicity events reveal the``ridge-like" structure, that is most prominent in the 1 to 2 GeV/c $p_{T}$-range, while it gradually decreases with increasing $p_{T}$-interval. 
The diminishing trend of correlated yield with the increasing $p_{T}$-interval is similar to what is observed in data. However, the EPOS3 events overestimate the correlated yields. The 
kinetic freeze-out temperature ($T_{kin}$) and a transverse radial flow velocity ($\beta$) at the freeze-out surface as obtained from the blast-wave fits to the identified particle spectra for 
high-multiplicity event classes of EPOS3-simulated events, do not match those parameters for the data. The mean transverse momentum as a function of multiplicity for all charged 
particle and as a function of mean multiplicity for identified charged particles from EPOS3 generated events, in different kinematic ranges, vary widely from those measured for 13 TeV 
pp data at LHC. The inverse slope parameters obtained by exponential fits to the transverse mass spectra of EPOS3 generated events, though exhibiting mass-ordering as expected 
in hydrodynamic model of particle production, the values of the parameter do not agree with those obtained from the data.\\

This data-driven study reveals that though the EPOS3 hydrodynamic model may reasonably describe the nature of some of the average bulk features, reflecting collective properties of 
particle production in the high-multiplicity pp events at the LHC, it cannot match the data quantitatively. With the quantitative mismatch in terms of the studied observables, related to the 
transverse momentum of the particles, one would expect a quantitative mismatch in terms of the fundamental observable, the inclusive spectra of invariant yields of produced particles, 
also. In this respect, the EPOS3 code appears inconsistent. The quantitative mismatch, however, cannot undermine the qualitative agreement of the EPOS3 simulation with the data. 
As has already been shown \cite {ref35} in a study on identified particle spectra data \cite {ref20} of pp collisions at 7 TeV, both the EPOS versions, EPOS3 and EPOS-LHC, match the 
trend of the data better than the conventional non-hydro event generators, including PYTHIA \cite {ref36}. The EPOS simulations, in contrast to PYTHIA, include flow of produced particles. 
The EPOS-LHC contains parameterized collective flow at the freeze-out and EPOS3, the advance version, contains a full (3D + 1) viscous hydrodynamic simulation. The better matching 
of the particle spectra data by EPOS simulations as compared to the conventional pp simulation codes establishes the collective behaviour of pp data at the LHC. Nevertheless, in the 
scenario when several other hydrodynamic as well as non-hydrodynamic models qualitatively reproduce the so-called collective features of particle productions in high-multiplicity small 
collision systems, quantitative descriptions of as many observables as possible in a consistent framework / model is desirable, to identify the origin of these features. This study may be 
useful in tuning the EPOS3 input parameters further, to match the high-multiplicity pp data. \\

In the context of this study, it is worth noting that some other hydro-based calculations have reported good agreement with a few observables of high-multiplicity events of small collision 
systems. Considering fluctuating proton IP-Glasma initial state model coupled to viscous hydrodynamic simulations, followed by the hadronic cascade model UrQMD, the harmonic flow 
coefficients of the pPb data at $\sqrt s_{NN} =$ 5.02 TeV could be well reproduced \cite {ref37}. In another study \cite {ref38}, in a viscous hydrodynamic model, the collective features of 
high-multiplicity events data of pPb collisions at LHC and (p,d,$^3$H)Au collisions at RHIC have been been quantitatively matched. A common hydrodynamic origin to the experimentally 
observed flow of produced particles has been suggested \cite {ref39} for all the large or the small central collision systems, including pp, pPb and PbPb at the LHC by considering a 
generalised initial condition calculations in the Glauber Model at the sub-nucleonic level, followed by a hybrid model that combines pre-equilibrium dynamics with viscous fluid dynamic 
evolution and late-stage hadronic re-scatterings. \\

On the other hand, some of the non-hydrodynamic models of particle productions also qualitatively match certain bulk collective features of high-multiplicity pp events. The Color Glass 
Condensate (CGC) effective theory also explains several of the experimental observations, including azimuthal correlations, of high-multiplicity small collision systems. In the IP-Glasma 
model, based on colour glass condensate, followed by the Lund string fragmentation algorithm of PYTHIA, with further tuning of the $p_{T}$-smearing fragmentation parameter in the 
default PYTHIA, the particle mass dependence of  $\langle p_{T} \rangle$ and the $p_{T}$ dependence of $v_{2}$ for pp collisions could be qualitatively reproduced \cite {ref40}. The 
literature also provide some other sources of long-range rapidity correlations, beyond hydrodynamic or CGC models, in terms of initial state physics processes. Several sources of such
long-range azimuthal correlations have been discussed in the review articles at references~ \cite {ref41} and \cite {ref42}. The ALICE measurement of $\langle p_{T} \rangle$ of the charged 
particles as a function of $N_{ch}$, including particles of $p_{T}$ up to 10 GeV/c from pp collisions at 7 TeV, could be well reproduced \cite {ref31} by invoking Colour Reconnection 
mechanism in PYTHIA (while EPOS3 remain far away from the data). The MPI model, however, cannot explain \cite {ref35} the dependence of $\langle p_{T} \rangle$ on $\langle 
N_{ch} \rangle$ or does not provide alternate explanations to other important features of particle production in high-multiplicity events pp collisions at 7 TeV when the $p_{T}$-range is 
restricted to the range of interest for studying the hydrodynamic collectivity. \\

The present study is consistent with ALICE study \cite {ref43} that observes the EPOS3 event generator is tuned to the LHC Run 1 data, to describe the inclusive transverse 
momentum spectrum for 13 TeV pp collisions reasonably well but not in detail. Similarly, no particle production model, hydrodynamic or non-hydrodynamic, could quantitatively match 
all the collective features of the high-multiplicity pp data, while the high-multiplicity proton-nucleus data have been better matched by several models \cite {ref37, ref38, ref39, ref40, ref44}, 
including EPOS3.  We, thus, conclude that the observed anomaly in particle production in high-multiplicity pp events at the LHC still remains unresolved, inviting further tuning of the existing 
models for quantitative description of the data.   \\
 
\section {Acknowledgement}
\label{}
The authors are thankful to Klaus Werner for providing them with the EPOS3 code. The authors also thank the members of the computing teams of the KANNAD of the C\&I Group of 
VECC for providing uninterrupted facility for event generation.\\

\end{document}